\documentclass[aps, prl,10pt,notitlepage,nofootinbib,superscriptaddress,showkeys,twocolumn]{revtex4}

\usepackage{hyperref}
\usepackage{amsfonts,amssymb,amsmath,exscale,bbm}
\usepackage[all,knot]{xy}  
\usepackage{color}
\usepackage{graphicx}
\usepackage{epsfig}
\usepackage{subfig}

\usepackage{amsmath}
\usepackage{amsfonts}
\usepackage{graphicx}
\usepackage{psfrag}
\usepackage{array}
\usepackage{bbm}
\usepackage{hyperref}
\usepackage{amsthm}

\theoremstyle{definition}

\newcommand\Tr{\mathrm{Tr}}
\newcommand\T{\mathrm{T}}

\newcommand{\cC}{{\mathcal C}}

\newcommand{\cE}{{\mathcal E}}
\newcommand{\cF}{{\mathcal F}}
\newcommand{\cG}{{\mathcal G}}

\newcommand{\cN}{{\mathcal N}}

\newcommand{\cT}{{\mathcal T}}
\newcommand{\cV}{{\mathcal V}}

\newcommand{\g}{{\mathfrak{g}}}

\newcommand{\R}{\mathbb{R}}

\newcommand{\Z}{\mathbb{Z}}

\newcommand\beq{\begin{equation}}
\newcommand\eeq{\end{equation}}
\newcommand{\be}{\begin{equation}}
\newcommand{\ee}{\end{equation}}
\newcommand{\bes}{\begin{eqnarray}}
\newcommand{\ees}{\end{eqnarray}}

\newcommand\beqa{\begin{eqnarray}}
\newcommand\eeqa{\end{eqnarray}}

\def\eps{\epsilon}

\def\vphi{{\varphi}}
\def\vphib{{\overline \varphi}}

\def\g{\mathfrak{g}}

\newcommand{\SU}{\mathrm{SU}}

\newcommand{\SO}{\mathrm{SO}}

\def\extd{\mathrm {d}}

\newcommand{\im}{\mathrm{im}}

\newcommand\acts\triangleright

\newcommand\maps{\colon}

\newcommand{\subjclass}[2][1991]{%
  \let\@oldtitle\@title%
  \gdef\@title{\@oldtitle\footnotetext{#1 \emph{Mathematics subject classification.} #2}}%
}

\begin{document}


\title{\bf 
Melonic phase transition in group field theory}

\author{{Aristide Baratin}}
\author{{Sylvain Carrozza}}
\author{{Daniele Oriti}}
\author{{James Ryan}}
\author{{Matteo Smerlak}}
\affiliation{Max-Planck-Institut f\"ur Gravitationsphysik, Am M\"uhlenberg 1, D-14476 Golm, Germany \\ email: name.surname@aei.mpg.de}


\begin{abstract}
Group field theories have recently been shown to admit a 1/N expansion dominated by so-called `melonic graphs', dual to triangulated spheres.    
In this note, we deepen the analysis of this melonic sector. We obtain a combinatorial formula for the melonic amplitudes in terms of a graph polynomial related to a higher dimensional generalization of the Kirchhoff tree-matrix theorem. Simple bounds on these amplitudes show the existence of a phase transition driven by melonic interaction processes. We restrict our study to the Boulatov-Ooguri models, which describe topological $BF$ theories and are the basis for the construction of four dimensional models of quantum gravity. 

\

\noindent {\footnotesize Mathematics subject classication: 81T25, 83C27  (principal), 83C45  (secondary).}

\end{abstract}


\maketitle


\section{Introduction}
\label{sec:intro}
Group field theory (GFT) is an approach to quantum gravity \cite{GFT} complementing loop quantum gravity \cite{LQG,Rovelli} and spin foam models \cite{SF}. Spin foams dynamically evolve the quantum states of geometry provided by loop quantum gravity; group field theory embeds them within a quantum field theoretic framework, in which spin foam complexes arise as Feynman diagrams \cite{Reisenberger:2000zc}. As a result, GFT provides a workable prescription for the quantum geometrodynamics of states in loop quantum gravity.
%
%
 
This field theoretic language allows for the application of key tools, absent from the other approaches. Two examples indicative of recent developments are an analysis of GFT renormalization \cite{GFTrenorm, GFTrenorm2,joseph}, and GFT symmetries \cite{GFTdiffeos,joseph2,joseph3}, as well as the development of new approximation schemes, within which to obtain effective cosmological dynamics \cite{GFTcosmo}.  These occur alongside the construction of interesting GFT models for 4d gravity \cite{SF,EPRL, aristidedaniele}.

Concurrently, group field theories are related to theories of random tensors, known as tensor models \cite{tensor}. This relationship is somewhat analogous to that between quantum field theory and quantum mechanics.    On the one hand, GFTs represent an enrichment of tensor models by group theoretic data -- this data is instrumental in providing a link to loop quantum gravity and various discrete gravity path integrals. On the other hand, tensor models form the backbone of the GFT formalism 
\cite{vincentTensor, GFTrenorm,GFTrenorm2}. Ultimately, the fact that both theories generate sums over cellular complexes, each of which represents a discrete counterpart to a continuum spacetime, means that similar statistical methods are applicable.

%
%
%
Such statistical analyses have already achieved remarkable success in explaining the content of these theories.  In the broader scheme, they can be best understood as an attempt to realize in higher dimensions those key steps leading to the quantization of $2d$ gravity via matrix models \cite{mm}.
 %
%
%
Thus, a $1/N$--expansion has recently been introduced in arbitrary dimensions,  dominated by so--called \lq\lq melonic graphs\rq\rq\ having spherical topology \cite{Gurau:2010ba, Gurau:2011xq, otherLargeN, scaling}.  
For a large class of tensor models, which include independent identically distributed (i.i.d.) \cite{critical, universal}, dually--weighted \cite{dariorazvan}, and matter coupled models \cite{ising, dimer, loop},  it has been demonstrated that there are critical values of the coupling constants indicative of a discrete--to--continuum transition \cite{uncoloring}. 
%
%
The existence and properties of such transitions should be investigated for fully--fledged group field theories, as they are the key to unraveling their non--perturbative features.

The present note is a step in this direction. It deepens the analysis of the leading order (or melonic) sector in the $1/N$--expansion of GFTs and consists of two main results:
\begin{description}
\item[1.]  We obtain a combinatorial formula for the melonic amplitudes in terms of a graph polynomial, by means of a higher--dimensional generalization of the Kirchhoff tree-matrix theorem. 
This reduces the evaluation of the amplitude to a counting problem, which will be a convenient starting point for the explicit computation of critical exponents. 

\item[2.]  Based on this formula, we obtain simple bounds on the amplitudes, which show the existence of a critical point where the partition function loses analyticity. 

\end{description}

We restrict our attention to Boulatov--Ooguri models \cite{boulatov, ooguri, GFT} in arbitrary dimensions. These describe topological $BF$ theories and are the basis for the construction of four-dimensional models of quantum gravity \cite{EPRL, aristidedaniele}. 



\section{Regularization and $1/N$ scaling}

We shall consider the colored version of Boulatov--Ooguri group field theories in any dimension $D$.  Given a compact connected Lie group $G$,\footnote{Typically $G=\SU(2)$ or $\SO(4)$.} the variables are a collection of $D+1$ complex tensor fields $\vphi_\ell \,:\,  G^{\times D} \to \mathbb{C}$, labeled by the color index $\ell\in\{0,\dots, D \}$. Each field satisfies a translation invariance under a diagonal action of the group:
\beq \label{eq:shift}
\vphi_\ell(hg_1, \cdots hg_D) = \vphi_\ell(g_1, \cdots g_D), \quad \forall h \in G. 
\eeq
The models considered here are characterized by a \lq\lq trivial\rq\rq\ kinetic term $\sum_\ell |\vphi_\ell|^2$ connecting each field $\vphi_\ell$ to its complex conjugate $\overline{\vphi}_\ell$:
\beq\label{eq:prop}
|\vphi_\ell|^2 = \int_{G^D} \prod_{i=1}^{D} \extd g_i \, \vphi_\ell(g_1 \cdots g_D) \overline{\vphi}_\ell(g_1, \cdots g_D)
\eeq
and by an interaction term of a specific simplicial type:
\beq\label{eq:int}
S_{int} (\vphi_\ell , \vphib_\ell) = \lambda \int \prod_{i<j} \extd g_{ij} \prod_{\ell = 0}^{D} \vphi_\ell ( {g_\ell}) \; + \; \rm{c.c.}
\eeq
where $\lambda$ is a coupling constant. We use the notation $g_\ell \equiv ( g_{\ell (\ell-1)} , \ldots , g_{\ell 0} , g_{\ell d} , \ldots , g_{\ell (\ell + 1)})$ and enforce the identification $g_{ij}=g_{ji}$. The complex conjugate term \lq\lq c.c.\rq\rq\ ensures that the action is real.
In the above formula,  $\extd g_i$ and $\extd g_{ij}$  denote the normalized Haar measure on $G$.

The form of the kinetic term, along with the symmetry \eqref{eq:shift}, means that the Feynman expansion is implemented with respect to a Gaussian measure $\extd \mu_{P}(\vphi^\ell,\vphib^\ell)$ with covariance $P$ (propagator):
\beq\label{eq:cov}
P(g_1, \cdots g_D; g'_1, \cdots g'_D) = \int_G \extd h \prod_i\delta (h g_i (g'_i)^{-1})
\eeq
The presence of  delta functions in the propagator leads to divergences in the Feynman graph amplitudes. 
The theory can be regularized by replacing these delta--functions with heat kernels \cite{Freidel:2004nb,matteovalentin1} at time $\tau$ on $G$,  given by:\footnote{Recall that $K_{\tau}$ is the solution of the heat equation on $G$:
\[(\partial_{\tau} - \Delta) K_{\tau} = 0\] 
with initial condition $\lim_{\tau\to 0} K_\tau(g) = \delta(g)$. $\Delta$ is the Laplacian on $G$.}
\beq \label{heat_kernel}
K_{\tau}(g) = \sum_\rho d_\rho\; e^{-\tau C_\rho}\;\Tr_{\rho}[g]\;,
\eeq
where $\Tr_{\rho}$ is the trace in the irreducible representation $\rho$ of $G$,  $d_\rho$ is its dimension and $C_\rho$ is the quadratic Casimir. For $G=\SU(2)$, $\rho\in\mathbb{N}/2$, $d_\rho = 2\rho+1$ and $C_\rho = \rho(\rho+1)$. 
We perform the following rescaling of the coupling constant:
\beq \label{scaling}
\lambda \quad \to \quad \lambda/{N_\tau^{(\dim G) \frac{(D-2)(D-1)}{4}}}\;,
\eeq
for some scaling function $N_\tau$.  This will be specified later to facilitate an interesting $1/N$ expansion in the $\tau \to 0$ limit. 

\begin{figure}[htb]
\centering
\includegraphics[scale=0.6]{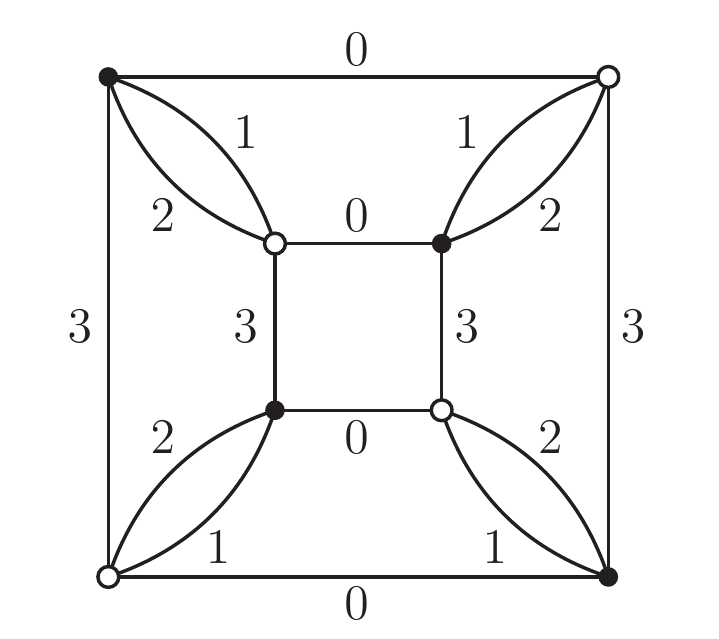}
\caption{\label{fig:colored}An example of a colored graph (with $D = 3$).}
\end{figure}

The Feynman expansion of the free energy of these models is a weighted sum over (closed connected) $(D+1)${\sl--colored graphs}\footnote{$(D+1)$--colored graphs are $(D+1)$--valent bipartite graphs, such that at any given vertex, the $D+1$ incident edges are labeled by distinct colors from the set $\{0,1,\dots, D\}$.} $\cG$ (see Figure \ref{fig:colored}):
\beq \label{fenergy}
F_{\tau, \lambda\overline\lambda} = \sum_\cG \frac{(\lambda\overline\lambda)^p}{\textsc{sym}(\cG)} A_\tau (\cG)
\eeq
where $A_\tau (\cG)$ is the graph amplitude and $\textsc{sym}(\cG)$ is a symmetry factor associated to $\cG$. 

The color information provides such a graph with the structure of a $D$-dimensional cellular complex\footnote{The $d$-cells  are defined as maximally connected subgraphs comprising of edges with $d$ fixed colors.} dual to a $D$-dimensional simplicial (pseudo)--manifold.

In what follows, we shall specifically consider the 2--complex $\cC_{\cG}$ associated to the graph $\cG$ comprising of its vertex, edge and face sets.\footnote{These correspond to the $0$--, $1$-- and $2$--cells of $\cG$, respectively.} These are denoted by  $\cV$, $\cE$ and $\cF$, respectively. 
For a given orientation of the faces, let $\epsilon_{fe} =\pm 1$ or $0$  be the face--edge adjacency matrix of size $|\cF|\times|\cE|$; it encodes the edge content of the faces and their relative orientations. The graph amplitude  $A_\tau (\cG)$ takes the form:
\beq \label{regamp}
A_\tau(\cG) =N_\tau^{-(\dim G) k_\cG} \int \prod_{e \in \cE} \extd h_e \prod_{f\in\cF} K_{m_f \tau} \left( \overrightarrow{ \prod_{e \in \partial f} } {h_e}^{\eps_{fe}}\right)
\eeq
where 
\beq
k_\cG =   \frac{(D-2)(D-1)}{4}|\cV|
\eeq 
and $m_f$ is the number of edges $e\in \partial f$ in the boundary of the face $f$. 
The appearance of the rescaled times $m_f \tau$ in (\ref{regamp}) is due to the $m_f$ iterations of the convolution property
of the heat kernel:
\beq
\int_G \extd g\, K_{\tau}(h g) K_{\tau'}(g^{-1}h') = K_{\tau+\tau'}(hh'), 
\eeq
in computing the face contribution of the amplitude (\ref{regamp}).  One can check this property directly from \eqref{heat_kernel} using the orthogonality relations for the representation matrices. 

Note that the amplitude enjoys a $G^{\times|\cV|}$ symmetry realised at the vertices of $\cG$,
\beq \label{gauge}
h_{e} \rightarrow k_{s(e)}\,h_e\,k_{t(e)}^{-1},
\eeq
where $s(e),t(e)\in \cV$ are the source and terminus of the oriented edge $e\in \cE$ and $k_v\in G$.  
One may remove this redundancy by fixing the value of the group elements along a spanning tree $\cT\subset \cE$ of the graph, say 
\beq \label{gaugefix}
h_e = 1, \quad \forall e\in \cT
\eeq
in the integrand of \eqref{regamp}, while dropping the corresponding integrals.  A residual symmetry at the root of $\cT$ contributes to a factor equaling the volume of the group with respect to the (normalized) Haar measure, namely $\textsc{vol}(G) = 1$. 
In the language of lattice gauge theory, the amplitude \eqref{regamp} integrates over discrete $G$--connections on $\cC_\cG$; 
the arguments of the heat kernels are the holonomies around the faces $f$;  (\ref{gauge}) is the discrete version of gauge invariance.

\begin{figure}[htb]
\centering
\includegraphics[scale=0.6]{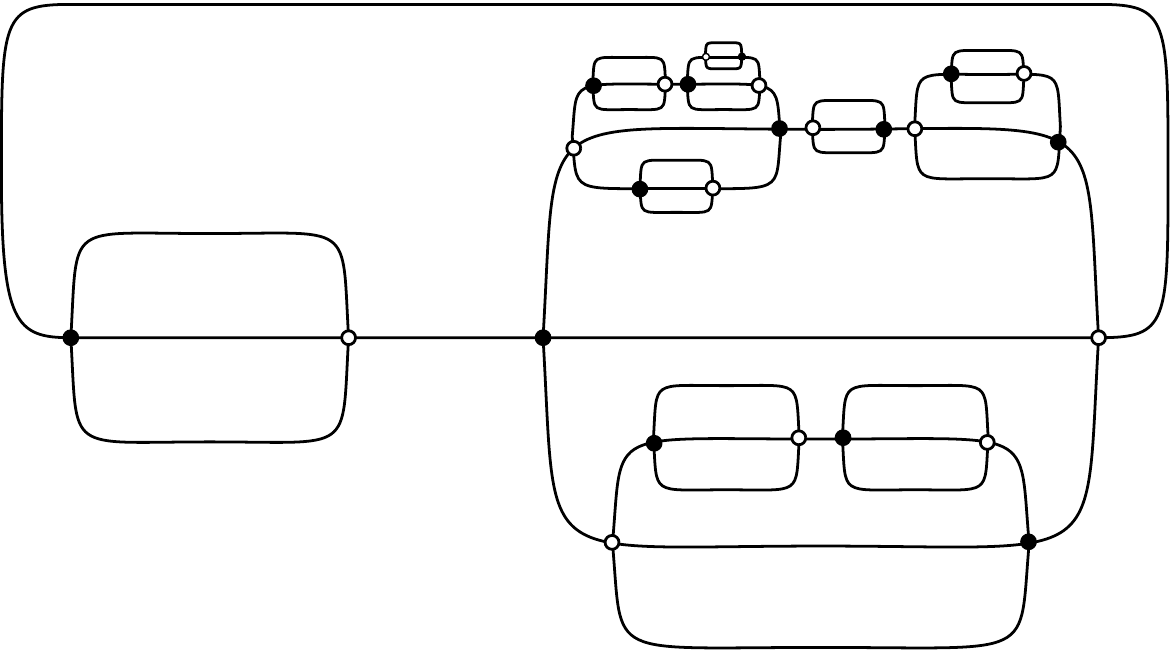}
\caption{\label{fig:melonic}An example of a melonic graph (with $D=3$).}
\end{figure}

As shown in \cite{Gurau:2011xq}, the choice 
\beq 
N_\tau = (4 \pi \tau)^{- \frac12}
\eeq 
for the scaling function (\ref{scaling}) allows one to organize the expansion (\ref{fenergy}) as: 
\beq
F_{\tau,\lambda\overline\lambda} \quad \underset{\tau \to 0}{\sim}\quad  N_\tau^{(\dim G)(D-1)}\; F_{\lambda\overline\lambda}^{(0)} \;,
\eeq
where the dominant contribution $F_{\lambda\overline\lambda}^{(0)}$ comes from the so--called {\sl melonic} subset of the graphs.  Given that such graphs have an iterative definition, let us denote by $M_p$ the set of melonic graphs with $2p$ vertices. Then:
\begin{description}
\item[$M_1$:] There is a unique such melonic graph, known as the supermelon, consisting of two vertices sharing $D+1$ edges. It is illustrated in Figure \ref{fig:super}

\begin{figure}[htb]
\centering
\includegraphics[scale=1]{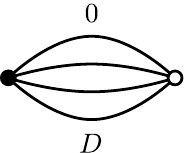}
\caption{\label{fig:super}The supermelon graph.}
\end{figure}

\item[$M_p$:] One generates the subset of melonic graphs $M_p$ from the subset $M_{p-1}$, by replacing an edge of a graph from $M_{p-1}$ with an instance of the partial subgraph, illustrated in Figure \ref{fig:elementary}.  This partial subgraph consists of two vertices sharing $D$ edges and is known as an elementary melon.

\begin{figure}[htb]
\centering
\includegraphics[scale=0.6]{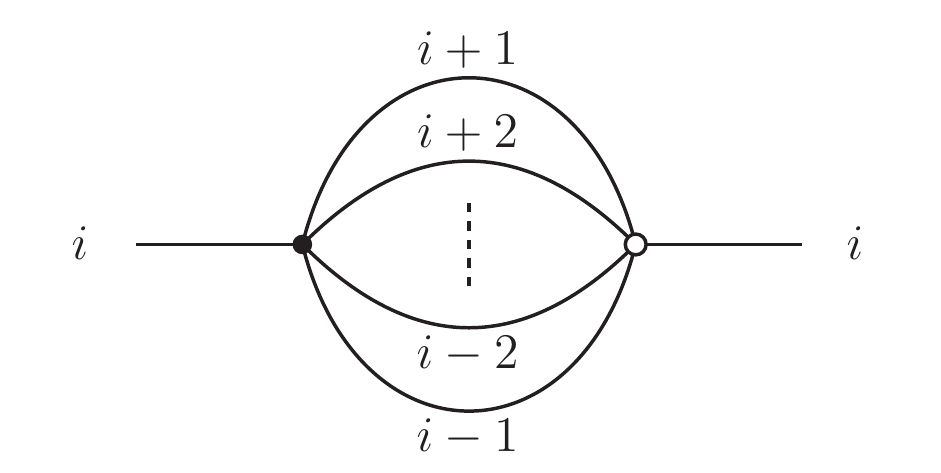}
\caption{\label{fig:elementary}The elementary melon of color $i$ to replace an edge of color $i$.}
\end{figure}

\end{description}
As a result, a generic melonic graph consists of elementary melons nested within elementary melons and so on. Such a graph is drawn in Figure \ref{fig:melonic}.

As a result:\footnote{For closed melonic graphs with $2p$ vertices, $\textsc{sym}(\cG) = p$.}
\beq
\label{eq:melonfree}
F_{\lambda\overline\lambda}^{(0)}  = \sum_{p \in \mathbb{N}} \frac{(\lambda \overline{\lambda})^p}{p} \sum_{\cG \in M_p} a(\cG)
\eeq
where
\beq \label{melonic}
a(\cG) = \lim_{\tau \to 0}\; N_\tau^{- (\dim G) (D-1)} A_\tau (\cG) < + \infty \;.
\eeq
In the next section, we analyse some properties of the coefficients $a(\cG)$ that allow us to determine the nature of the dominant series $\cF_{\lambda\overline\lambda}^{(0)}$ as a function of $\lambda\overline\lambda$.

\section{
Melonic amplitudes}


We want to take the limit $\tau \to 0$ in the formula \eqref{melonic}. 
The first step is to use the standard \cite{camporesi} small-times asymptotics of the heat kernels:\footnote{More explicitly, there exists an asymptotic expansion for the heat kernel, valid for small $\tau$, the first term of which is:
\begin{equation*}
K_\tau(g) \underset{\tau\rightarrow 0}{\sim} (4\pi \tau)^{-\frac{\dim G}{2}} e^{-\frac{|g|^2}{4\tau}} \sqrt{D_{\textsc{vvm}}(g)}\;,
\end{equation*} 
where $D_{\textsc{vvm}}(g) = |g|^{-1/2}\vert\det(\partial_\mu\partial_\nu|g|^2/2)\vert$ is the Van Vleck--Morette determinant \cite{camporesi}.  In the small $\tau\to 0$ limit, this evaluates to a constant and so does not affect our subsequent analysis. Moreover, $|g|$ is the geodesic distance of $g$ from the identity of the group evaluated using the bi--invariant metric on the group. Replacing the group element $g$ by its algebra element $X$ gives rise to the Killing form with the normalization factors quoted in \eqref{eq:heatasymp}.
 }
\beq
\label{eq:heatasymp}
K_{\tau}(e^X) \underset{\tau\rightarrow 0}{\sim} (4\pi \tau)^{-\frac{\dim G}{2}} e^{-\frac{\langle X, X\rangle}{4\tau}}, 
\eeq
to recast the amplitudes $A_{\tau}(\cG)$ as Gaussian integrals. 
Here we wrote group elements as exponentials of Lie algebra elements $X\in \g$; $\langle\cdot,\cdot\rangle$ is the Killing form on $\mathfrak{g}$. 

The second step is to evaluate the resulting integrals by Laplace's method \cite{matteovalentin1}. 
In the limit $\tau \to 0$, the integral  occurring in (\ref{regamp})  is dominated by {\sl flat} discrete connections. 
Since the 2--complex $\cC_\cG$ is simply connected, the unique flat connection (after gauge fixing along a tree $\cT$) is the trivial one:
$h_e \!=\! 1, \forall e\in \cE$. 
At the neighborhood of the identity, one can trade the group integrals occurring in (\ref{regamp}) by integrals over Lie algebra elements $X_e \in \mathfrak{g}$. We thus obtain:\footnote{We refrained from including a constant Jacobian factor contributed by the change from the Haar measure on the group to the Lebesgue measure on the algebra.}
\beq \label{gauss}
A_\tau(\cG)\underset{\tau\rightarrow 0}{\sim} \left[\frac{N_\tau^{2(  |\cF| - k_\cG)}}{\prod_{f} m_f }\right]^{\frac{\dim G}{2}} \hspace{-3pt} \int \prod_{e\in \tilde \cE } \extd X_e\,  e^{ - \frac{S(X_e)}{4 \tau}} 
\eeq
Here $\extd X_e$ is the Lebesgue measure on $\mathfrak{g} \simeq \R^{\dim G}$, 
the product is over all edges in $\widetilde \cE = \cE \backslash \cT$, and 
$S(X_e)$ is the quadratic form given by:
\beq
S(X_e) = \sum_{f \in \cF} \frac{1}{m_f}\sum_{\substack{e, e'\in \partial f\backslash\cT }} \epsilon_{fe} \epsilon_{fe'} \langle X_e,  X_{e'} \rangle
\eeq
We consider the symmetric $|\cE| \times |\cE|$ matrix $L = \epsilon^\intercal D_m \epsilon$, where $D_m$ is the diagonal matrix with entries
$(D_m)_{ff} = 1/ m_f$ indexed by $\cF$, and $\epsilon^\intercal_{ef}:= \epsilon_{fe}$ is the transpose adjacency matrix. 
In the terminology of homology, $\epsilon^\intercal$ is  the boundary map  $\partial_2$ from faces to edges of the  2-complex $\cC_{\cG}$ and  $L$ is the second (weighted) Laplacian matrix, with entries: 
\beq
L_{e,e'} = \sum_f \frac{1}{m_f} (\partial_2)_{ef} (\partial_2)_{fe'}^\intercal
\eeq 
We also denote by $\widetilde{L}$ the submatrix of $L$ with rows and columns indexed by $\widetilde\cE$. 
Upon Gaussian integration in (\ref{gauss}), we obtain:  
\beq
A_\tau(\cG) \underset{\tau\rightarrow 0}{\sim} \left[\frac{N_\tau^{2(|\cF|  - |\widetilde\cE| - k_\cG)}}{\det (\widetilde{L}) \prod_{f} m_f 
 }\right]^{\frac{\dim G}{2}}
\eeq

The third step is to observe that for melonic graphs:
\begin{equation}
\label{eq:melonRel}
|\cE| =|\cV|\frac{(D+1)}{2}\,, \quad |\cF| = D +|\cV| \frac{D(D-1)}{4}
\end{equation}
Together with the relation\footnote{Recall that a spaninng tree $\cT$ has $|\cV|-1$ edges.} $|\widetilde\cE| = |\cE| - |\cV|+1$,  
it yields:
\beq
|\cF|  - |\widetilde\cE| - k_\cG = D-1
\eeq
so that the dependance upon $N_{\tau}$ drops in the limit (\ref{melonic}). We obtain the following expression for the melonic amplitudes: 
\begin{equation}
\label{eq:AGlaplace}
a(\cG) =    \left[\det (\widetilde L) \prod_{f} m_f\right]^{-\frac{\dim G}{2}} \,.
\end{equation}

\section{Tree expansion}

In this section,  we dwell upon the determinant showing up in expression \eqref{eq:AGlaplace} for the melonic amplitudes. 
As we shall see, just as the Kirchhoff tree--matrix theorem expresses the determinant of the first Laplacian $\partial_1 \partial_1^\intercal$ of a graph as a sum over all its spanning trees, one may expand the determinant of the second Laplacian of a 2--complex over two--dimensional analogues of trees \cite{simplicial}. 
Such an expansion shows that the problem of estimating melonic amplitudes is equivalent to the purely combinatorial problem of enumerating  such \lq\lq 2--trees\rq\rq, on which several results exist in the combinatorics literature \cite{adin1992counting, peterssonenumeration, simplicial}. 

The tree expansion formula goes as follows:
\beq \label{Texpansion}
\det \widetilde L = \sum_{\mathrm{\cC_\T} \in \cT_{2}(\cG)} \! |H_{1}(\cC_\T)|^2 \prod_{f\in \cC_\T} \frac{1}{m_f}
\eeq 
The sum runs over all {\sl spanning 2--tree}s of the 2--complex $\cC_\cG$, defined as subcomplexes  $\cC_\T \subset \cC_\cG$ containing all edges and all vertices of $\cG$ and such that the following conditions hold: 
\beq \label{acyclic}
H_2(\cC_\T) = 0 \quad \mbox{and} \quad |H_1(\cC_\T)| < \infty
\eeq
where $H_i(\cC_\T)$ denotes the $i$-th (integral) homology group\footnote{We are using the standard cellular homology.
Given a cell complex $X$, we consider the chain groups $C_i(X)$, namely the free abelian group generated by the $i$-dimensional faces, and the standard boundary operators:
\[
\partial_i\maps C_i(X) \to C_{i-1}(X)
\]
The $i$-th homology group is $H_i(X)=\ker \partial_i/\im\, \partial _{i+1}$.}  
of $\cC_\T$ and $|H_i|$ is the cardinality of $H_i$.  The fact that $H_1(\cC_\T)$ is a finite group simply says that $\cC_\T$ is simply connected. 
Intuitively, $\cC_\T$ has enough faces to keep all loops of edges contractible, but not too many, so that they do not form higher dimensional cycles.

The main ingredient to prove the expansion (\ref{Texpansion}) is the Cauchy--Binet identity for the determinant of square matrices composed of non--square factors. Our matrix $\widetilde{L}$ takes the form of a product $A^\intercal B$, where $A, B$ are matrices with rows and columns indexed respectively by faces $f \in \cF$ and edges $e\in \widetilde\cE$. The Cauchy--Binet formula expands the determinant over subsets $\T \subset \cF$ of cardinality $|\widetilde{\cE}|$:
\beq \label{CB}
\det \widetilde{L} = \sum_{\substack{\T \subset \cF \\|\T|=|\widetilde\cE|}} |\det \partial_{2 | \widetilde\cE, \T}|^2 \det D_{m|\T}
\eeq
where $\partial_{2 | \widetilde\cE, \T}$ is the square submatrix of the second boundary map $\partial_2$ with rows and columns indexed respectively by $\widetilde\cE$ and $\T$,  and $D_{m|\T}$ is the submatrix of $D_m$ with rows and columns indexed by $\T$.

Given $\T\subset \cG$, we denote by  $\cC_{\T}$ the 2-complex having T as face sets and  containing all edges and all vertices of $\cG$. Because of the factor $|\det \partial_{2 | \widetilde\cE, \T}|$, the Cauchy-Binet sum is actually indexed by those subsets $\T$ such that $\partial_{2 | \widetilde\cE, \T}$ is non-degenerate.  The next step is to show that such subsets are precisely the face sets of 2--trees $\cC_\T \in \cT_2(\cG)$. 

Doing so requires the machinery  of relative homology\footnote{Given a subcomplex $A\subset X$ of a cell complex $X$,  we may consider the groups of relative chains, defined the quotients:  
\[C_i(X, A) = C_i(X)/C_i(A)\] of the chain groups.  
The homology of $X$ relative to $A$ is then the homology of the induced boundary maps: 
\[ 
\tilde\partial_i\maps C_i (X, A)\to C_{i-1}(X, A).
\]
namely:
\[
H_i(X, A) = \ker \tilde\partial_i/\im \tilde\partial _{i+1}
\]
A fundamental property of the relative homology groups is that they fit into the exact sequence of groups:
\beq \label{relative-hom}
\cdots \to H_n(A) \to H_n(X) \to H_n(X,A) \to H_{n-1}(A) \to \cdots
\eeq}.
Since the 2-complex $\cC_\T$ contains all edges of $\cG$, it contains the gauge--fixing tree $\cT$ as a 1-dimensional subcomplex, and we may consider the homology of $\cC_\T$ relative to $\cT$. 
$\partial_{2 | \widetilde\cE, \T}$ is then the matrix of the induced boundary map, so that its kernel defines the second relative homology group:
\beq 
\mathrm{ker} \,\partial_{2 | \widetilde\cE, \T} = H_2(\cC_\T, \cT)\;.
\eeq 
Therefore, the matrix is non-degenerate if and only if $H_2(\cC_\T, \cT) =0$. In virtue of the property (\ref{relative-hom}),  the relative homology group fits into the exact sequence:
\beq \label{sequence}
0 \to H_2(\cC_\T) \to H_2(\cC_\T, \cT)\to H_1(\cT) \to \cdots 
\eeq 
However,  $\cT$ is a tree.\footnote{Since $\cT$ has no loop, $H_1(\cT)$ is finite; but
as a top--dimensional homology group, it is also torsion--free, so $H_1(\cT)=0$.} As a result, $H_1(\cT) = 0$, so (\ref{sequence}) implies:
\beq
H_2(\cC_\T, \cT) \simeq  H_2(\cC_\T) \;.
\eeq
Therefore, the Cauchy--Binet sum is indexed by subsets $\T$ such that $H_2(\cC_\T)=0$, which is the first defining property of a 2--tree. 
The second  one arises from a cardinality argument, making use of the Euler--Poincar\'e identity. 
Given $\T \subset \cF$, the Euler characteristic of the 2--complex $\cC_{\T}$ is given by:
\beq 
\chi(\cC_\T) = |\cV|- |\cE| +  |\T|\;.
\eeq 
This can also be written as an alternating sum of Betti numbers.\footnote{$\chi(\cC_\T) = \beta_0(\cC_\T) -\beta_1(\cC_\T) + \beta_2(\cC_\T)\;.$} $\cC_\T$ is connected, so  $\beta_0(\cC_\T) = 1$; also if $H_2(\cC_\cT) = 0$, 
then $\beta_2(\cC_\cT) = 0$. Consequently:
\beq
\chi(\cC_\T) = 1 - \beta_1(\cC_\T) \;.
\eeq 
Considering that $|\widetilde\cE| = |\cE|- |\cV| +1$, equating the above two equalities yields:
\beq
|\T| - |\widetilde\cE| +  \beta_1(\cC_\T)= 0 \;.
\eeq
We conclude that when $H_2(\cC_\T)=0$, then $|\T| =|\widetilde\cE|$ if and only if $\beta_1(\cC_\T) = 0$, i.e  $H_1(\cC_\T)$ is finite. This gives the second defining property of a 2--tree. 
 
The last step is to show that $|\det \partial_{2 | \widetilde\cE, \T}| = |H_1(\cC_\T)|$ for any 2--tree $\cC_\T \in \cT_2(\cG)$. 
Once again, we shall make use of relative homology. 
First, since the tree $\cT$ is connected and simply connected,  the homology group $H_1(\cC_\T)$ fits into the exact sequence:
\beq
0 
\to H_1(\cC_\T) \to H_1(\cC_\T, \cT) \to 
0
\eeq
which says that $H_1(\cC_\T) \simeq H_1(\cC_\T, \cT)$. Second, since the complexes $\cC_\T$ and $\cT$ have the same vertices, the first induced boundary map $\widetilde\partial_1$ is the zero map, i.e $\ker \widetilde\partial_1 \! =\! \Z^{|\widetilde\cE|}$.  The first relative homology group is by definition the quotient of this kernel with  the image  of the second induced boundary map $\Z^{|\T|} \to \Z^{|\widetilde\cE|}$, so we get: 
\beq
H_1(\cC_\T, \cT) \simeq \Z^{|\widetilde\cE|} /  \partial_{2 | \widetilde\cE, \T}(\Z^{|\T|})
\eeq
The following lemma brings our argument to a conclusion:  \textit{Given a matrix $M$ of size $n \times n$ with integer coefficients and $M(\Z^n)$ the image of $\Z^n$ by the corresponding linear map, then if $M$ is non-degenerate the quotient group $\Z^n/M(\Z^n)$ is finite and its cardinality is $|\det M|$.}\footnote{This can be seen by using the \lq\lq Schmidt normal form\rq\rq\ decomposition of integer matrices: $M=UDV$ for some $U,V \in \rm{GL}(\Z)$ (invertible over the integers) and a diagonal matrix $D=\rm{diag}(d_1, .. d_n)$. $U(\Z^n) = \Z^n$ and $V(\Z^n)=\Z^n$, so $M(\Z^n) = D(\Z^n)$ and the quotient $Q=\Z^n/M(\Z^n)$ is the just the direct product of all $\Z/ d_i \Z$. Therefore $|Q| = \prod_i |d_i| = |\det D| = |\det M|$.}  
Applying this lemma to $\partial_{2 | \widetilde\cE, \T}$ gives the desired result. 

We have thus shown that the identity (\ref{CB})  reproduces the expansion (\ref{Texpansion}). At this stage we have our first result:

\vspace{0.2cm}

\noindent{\bf 1st Result:}
{\it The melonic amplitudes $a(\cG)$ may be expressed as weighted sums over their spanning 2--trees:
\begin{equation} \label{final}
a(\cG)
= \left[\sum_{\mathrm{\cC_\T} \in \cT_{2}(\cG)} \! |H_{1}(\cC_\T)|^2 \prod_{f \notin \cC_\T} m_f \right]^{- \frac{\dim G}{2}} \;.
\end{equation}
}
\vspace{0.2cm}

Let us close this section by looking specifically at the 3--dimensional case, where the group field theory \eqref{eq:int} gives a model for three--dimensional gravity \cite{boulatov}. Interestingly, in this case, there is a one--to--one correspondence between spanning 2--trees of $\cC_\cG$ and spanning trees of lines in the simplicial complex $\Delta_{\cG}$ dual to $\cC_\cG$. More precisely, via the map between the graph $\cG$ and its dual simplicial complex $\Delta_\cG$, the face sets $\T$ of 2--trees $\cC_{\cT}$ in the 2--complex correspond to edge sets $\widetilde \cE^* = \cE^* \setminus {\cT^*}$ where $\cE^*$ is the set of edges of $\Delta_\cG$ and ${\cT^*}$ is a spanning tree of edges in $\Delta_\cG$.\footnote{This identification stems from the fact that the second boundary map ${\partial}^*_2$ of $\Delta_{\cG}$ is just the transpose $\partial_2^\intercal$ of the second boundary map in $\cC_\cG$.   Just as the non--degeneracy of the submatrix $\partial_{2 | \widetilde\cE, \T}$ with $\widetilde{\cE} = \cE\backslash \cT$ characterizes spanning trees $\cT$ and 2--trees $\cC_\T$ in $\cC_\cG$, the non--degeneracy of $\tilde{\partial}_{2 | \widetilde\cE^*, \T^*}$
characterizes spanning trees ${\cT^*}$ and 2--trees $\cC_{\T^*}$ in $\Delta_{\cG}$.
This shows that spanning 2--trees in one complex are dual to spanning trees in the other.}
Therefore when $D=3$,  the melonic amplitudes can be written as a sum over spanning trees of the dual triangulation.  Spanning trees of the dual triangulation also play a key role for the gauge-fixing of the shift symmetry in 3-dimensional spin foam gravity \cite{Freidel:2004vi}.

There is a further simplification in the 3-dimensional case. By a (tedious, but straighforward) recursion on the number of vertices of  $\cG$, 
one can prove that every $k\times k$ minor of the boundary map $\partial_{2}$, where $k=\rm{rank}(\partial_2)$, equals 0 or $\pm 1$.  
 In particular the first homology group  of any 2--tree is trivial\footnote{This says that no \lq\lq torsionful 2--tree\rq\rq\ exists for 3--dimensional melonic graphs. It would be interesting to find a general homological argument for this fact and to see whether it holds true also in higher dimensions.} 
 and the homology factor disappears from the expression \eqref{final}. As a result, the formula for $a(\cG)$ is in this case, with $G=\SU(2)$:
\beq
a_{3\mathrm{D}}(\cG) = \left[ \sum_{{\cT^*} \in \cT_1(\Delta_{\cG})}  \prod_{l \in \cT^*} m_{f(l)} \right]^{- \frac{3}{2}} \;,
\eeq
where $f(l)$ denotes the face of $\cC_\cG$ dual to the link $l$ and $\cT_1(\Delta_{\cG})$ denotes the set of trees in the simplicial complex $\Delta_{\cG}$. 
We recognize the Kirchhoff polynomial for the (1-skeleton graph of) the simplicial complex.

\section{Bounds and nature of the series}

To show the presence of a phase transition, we now bound the radius of convergence of the series \eqref{eq:melonfree}. This bounding procedure is divided into three steps:
\begin{description}
\item[i: Bounds on $|\det(\partial_{2|\widetilde\cE,\T})|^2$.] The matrix has at most  $D$ non-trivial coefficients per line,
which means that the column vectors of $\partial_{2|\widetilde\cE,\T}$ have a Euclidean norm smaller than $D$. Therefore, by Hadamard's bound, we conclude that:\footnote{For $(D+1)$--colored graphs: $|\cE| = \frac{D+1}{2}|\cV|$.}
\beq
(1\leq\,) |\det(\partial_{2|\widetilde\cE,\T})|^2 \leq D^{|\widetilde\cE|} = D^{(D-1) p + 1}\;.
\eeq

\item[ii: Bounds on $a(\cG_p )$.] For each $2$--tree $\cC_\T$, we can easily bound $\prod_{f \notin \cC_\T} m_f$ by a geometric series in $p$ (for instance using the arithmetico--geometric inequality):
\bes
1 &\leq& \prod_{f \notin \cC_\T} m_f\\ &\leq& \left( \frac{1}{|\cF| - |\T|} \sum_{f \notin \cC_\T} m_f \right)^{|\cF| - |\T|} \\
&\leq& \left( \frac{D |\cE|}{|\cF| - |\T|} \right)^{|\cF| - |\T|} \\
&=& \left( \frac{D (D +1)p}{(D-1) + \frac{(D-1)(D-2)}{2} p} \right)^{(D-1) + \frac{(D-1)(D-2)}{2} p} \\[0.2cm]
&\leq& k_1 {c_1}^{p} 
\ees
for some constants $k_1 > 0$ and $c_1 > 0$. Therefore, calling $\cN(\cG_p)$ the number of $2$--trees in $\cG_p$, one has:
\beq
\left( k_1 {c_1}^{p} \cN(\cG_p) \right)^{\frac{- D}{2}} \leq a(\cG_p ) \leq \cN(\cG_p)^{-\frac{D}{2}}
\eeq

\item[iii: Bounds on the series.] The number of $2$--trees in a melonic graph of order $p$ is trivially bounded by $2^{|\cF|} = k_3 {c_3}^p$.\footnote{For a melonic graph: $|\cF| = \frac{D(D+1)}{4}|\cV| + D$.} Therefore, there exists constants $k > 0$ and $c > 0$ such that:
\beq
k \sum_{p \in \mathbb{N}} F_p c^p (\lambda\bar\lambda)^p \leq F^{(0)}_{\lambda\bar\lambda} \leq \sum_{p\in\mathbb{N}}F_p (\lambda\bar\lambda)^p\;.
\eeq 
where $F_p = \frac{1}{(D+1)p +1} {(D+1)p +1 \choose p}$ is an exact counting of the number of melonic graphs with $2p$ vertices.
As a matter of fact, the upper bound given here $\sum_{p\in\mathbb{N}}F_p (\lambda\bar\lambda)^p =  F^{(0)}_{i.i.d.}$ is the melonic free energy of the i.i.d.\ tensor model in $D$ dimensions.

\end{description}

Since $F^{(0)}_{i.i.d.}$ has a finite radius of convergence, so does the series with coefficients $\cF_p c^p g^p$, and we can conclude that $F^{(0)}_{\lambda\bar\lambda}$ itself has a finite radius of convergence.   Thus, we have our second result:

\vspace{0.2cm}

\noindent{\bf 2nd result:} {\it At leading order in the $1/N$--expansion, the free energy of topological group field theories possesses critical behaviour.}

\vspace{0.2cm}

Specialising again to three dimensions:

\vspace{0.2cm}

\noindent{\bf 2nd${}^\prime$ result:} {\it At leading order in the $1/N$--expansion, the free energy of the group field theory for 3--dimensional quantum gravity possesses critical behaviour.}

\vspace{0.2cm}

The present analysis leaves open a number of interesting questions for future study. In particular, one would like to gather information about the behaviour of $\cF^{(0)}_{\lambda\bar\lambda}$ close to its critical point, including its critical exponent. Other aspects of the ``tensor track'' \cite{vincentTensor} applied to GFT will include the extension of our results to non-melonic contributions \cite{Bonzom:2012wa} and the definition of a double scaling limit \cite{Gurau:2011sk,Kaminski:2013maa}. Finally, it is most important to extend the analysis to GFT models of four dimensional quantum gravity and to study the existence and properties of analogous phase transitions. 
Investigating the physical nature of such transitions will shed light on the continuum geometric properties of GFTs, and hence also of loop quantum gravity and its covariant spin foam formulation.



\vspace{0.5cm}

~\\
{\bf \large Acknowledgements}
~\\

We thank Frank Hellmann for helpful discussions at the beginning of this project.  We gratefully acknowledge the support of  the A.\ von Humboldt Stiftung through a Sofja Kovalevskaja Award.

\vspace{0.5cm}
\bibliographystyle{hunsrt}

\end{document}